\documentclass[%
preprint,
superscriptaddress,
pra,
]{revtex4}

\usepackage{amsmath}
\usepackage{mathrsfs}
\usepackage{amssymb}

\usepackage {xcolor}
\usepackage{comment}

\usepackage{subfigure}
\usepackage[colorlinks,
            linkcolor=blue,
            anchorcolor=blue,
            citecolor=blue]{hyperref}
\newtheorem{theorem}{Theorem}

\def\ra{\rangle}
\def\la{\langle}

\usepackage{graphicx}
\usepackage{dcolumn}
\usepackage{bm}

\begin{document}

\title{Entanglement certification from moments of positive maps}

\author{Qing-Hua Zhang}
\affiliation{School of Mathematics and Statistics, Changsha University of Science and Technology, Changsha 410114, China}

\author{Xiaoyu Ma}
\email[]{xyu.ma@outlook.com}
\affiliation{College of Science, National University of Defense Technology, Changsha 410073, China}

\author{Shao-Ming Fei}
\affiliation{School of Mathematical Sciences, Capital Normal University,
Beijing 100048, China}

\begin{abstract}
Entanglement certification is crucial in physical experiments, particularly when only partial knowledge of the quantum state is available. In this context, we present an entanglement criterion based on positive but not completely positive maps, which eliminates the need to identify eigenvalues of the output state. Notably, the Faddeev-LeVerrier algorithm establishes a relationship between the coefficients of characteristic polynomials and the moments of a matrix. This enables the existence of negative eigenvalues through the moments of the output state. The effectiveness of our criterion relies on the selection of positive maps, similar to the original positive maps criterion.
\end{abstract}

\maketitle

\section{Introduction}

Einstein, Podolsky and Rosen (EPR) and Schr\"odinger first recognized a ``spook" feature of quantum mechanics \cite{PhysRev.47.777,schrodinger1935gegenwartige}. Later, Bell observed that entanglement leds to experimentally testable deviations from classical physics to quantum mechanics \cite{PhysicsPhysiqueFizika.1.195}. In the framework  of quantum resource theory, entanglement has been identified as a resource with crucial applications in quantum information processing such as quantum teleportation \cite{PhysRevLett.70.1895}, quantum cryptography \cite{PhysRevLett.67.661} and measurement based quantum computation \cite{PhysRevLett.86.5188}. A fundamental problem in this context is the detection and quantification of quantum entanglement. Unfortunately, determining whether a given quantum state is entangled or separable remains an open problem \cite{RevModPhys.81.865,GUHNE20091}.

Up to now, many kinds of entanglement criteria have been proposed for the detection of entanglement in arbitrary quantum systems \cite{RevModPhys.81.865}. One of the most notable criteria for detecting entanglement is the positive partial transpose (PPT) criterion proposed by Peres-Horodecki \cite{PhysRevLett.77.1413,PhysRevLett.82.1056,HORODECKI19961}. Another significant criterion for detecting bound entanglement is the computable cross norm \cite{rudolph2005further} or realignment \cite{Chen2002Quantum} (CCNR) criterion.

Recently, Elben $et\ al.$ \cite{PhysRevLett.125.200501} proposed a method for detecting bipartite entanglement based on estimating moments of the partially transposed density matrix. For all $d_1 \otimes d_2$ dimensional separable bipartite states $\rho$ in systems $A$ and $B$, the $p_3$-PPT condition says that
\begin{equation}\label{p3ppt}
L_1 :\ \left[p_2\left(\rho^{T_B}\right)\right]^2-p_3\left(\rho^{T_B}\right) p_1\left(\rho^{T_B}\right) \leqslant 0,
\end{equation}
where $T_B$ denotes the partial transposition on the subsystem $B$, the $k$th partial moment is defined as $p_k\left(\rho^{T_B}\right)=\operatorname{Tr}\left[\left(\rho^{T_B}\right)^k\right]$. The violation of the inequality (\ref{p3ppt}) indicates that the state $\rho$ is an non-positive partial transpose (NPT) entangled state. Later, Neven $et\ al.$ \cite{neven2021symmetry} proposed a set of inequalities, known as $D_k^{(\text {in) }}$ inequalities, to detect bipartite NPT entangled states. The first nontrivial condition from the $D_3^{(\mathrm{in})}$ inequality reads
\begin{equation}
L_2 :\ \frac{3}{2}\left[p_1\left(\rho^{T_B}\right)\right]\left[p_2\left(\rho^{T_B}\right)\right]
-\frac{1}{2}\left[p_1\left(\rho^{T_B}\right)\right]^3-p_3\left(\rho^{T_B}\right) \leqslant 0
\end{equation}
for all separable states. When $1 / 2 \leqslant$ $p_2\left(\rho^{T_B}\right) \leqslant 1$, the $D_3^{(\mathrm{in})}$ criterion detects more entangled states than the $p_3$-PPT criterion, while the $p_3$-PPT criterion detects more entangled states than the $D_3^{\text {(in) }}$ criterion for $0 \leqslant p_2\left(\rho^{T_B}\right)<1 / 2$ \cite{neven2021symmetry}.

Yu $et\ al.$ \cite{PhysRevLett.127.060504} introduced an optimal entanglement detection criterion based on partial moments called the $p_3$-OPPT criterion. If $\rho$ is separable, the following inequality holds:
\begin{equation}
L_3:\ \mu x^3+(1-\mu x)^3-p_3\left(\rho^{T_B}\right) \leqslant 0,
\end{equation}
where $x=\frac{\mu+\sqrt{\mu\left[p_2\left(\rho^{T_B}\right)(\mu+1)-1\right]}}{\mu(\mu+1)}$ and $\mu=\frac{1}{p_2\left(\rho^{T_B}\right)}$. Another separability criterion based on Hankel matrices has been also derived in Ref.~\cite{PhysRevLett.127.060504}. For $q=\left(p_0,\ p_1,\ p_2,\ \cdots,\ p_n\right)$, the Hankel matrices are constructed by $\left[B_l(q)\right]_{i j}=p_{i+j+1}$ for $i, j=$ $0,\ 1,\ 2, \cdots, l$. If $\rho$ is separable, then for $l=1,\ 2,\ \cdots,\ \left\lfloor\frac{n-1}{2}\right\rfloor$, one has
$$
\begin{aligned}
L_4:\ & \widehat{B}_l(q)=\left[p_{i+j+1}\left(\rho^{T_B}\right)\right] \geqslant 0
\end{aligned}
$$
with $p_1\left(\rho^{T_B}\right)=1$. Later, Wang $et\ al.$ generalized the method based on partial transposition to the one based on positive but not completely positive maps \cite{wang2022operational}.

Zhang $et\ al.$ \cite{zhang2022quantum} proposed another entanglement detection criterion in terms of the so called realignment moments.
For a given $m \times n$ matrix $C=\left[c_{i j}\right]$ with entries $c_{i j}$, the vector $\operatorname{vec}(C)$ is defined by
$
\operatorname{vec}(C)=\left(c_{11},\ \cdots,\ c_{1 n},\ c_{21},\ \cdots,\ c_{2 n},\ \cdots,\ c_{m 1},\ \cdots,\ c_{m n}\right)
$. Let $M$ be an $m \times m$ block matrix with $n \times n$ block matrices. The realigned matrix $M^R$ is defined by
$
M^R=(\operatorname{vec}\left(M_{1,1}\right)^T,\cdots,\operatorname{vec}\left(M_{1, m}\right)^T,\cdots,\operatorname{vec}\left(M_{m, 1}\right)^T,\cdots,
\operatorname{vec}\left(M_{m, m}\right)^T)^T.
$
For a $d \otimes d$ dimensional bipartite state $\rho$, the realignment moments are defined by
$$
r_k\left(\rho^R\right)=\operatorname{Tr}\left[\left(\left(\rho^R\right)^{\dagger}\rho^R\right)^{k / 2}\right],~~ k=1,2, \cdots, d^2.
$$
If a quantum state $\rho$ is separable, then
\begin{equation}\label{zhang}
L_5:\ \left[r_2\left(\rho^R\right)\right]^2-r_3\left(\rho^R\right) \leqslant 0 .
\end{equation}

In \cite{PhysRevA.109.012404}, Aggarwal $et\ al.$ proposed a so-called R moments criterion by using the moments of the realigned density matrix. The approach enables the detection of entanglement for both distillable and bound entangled states. For all separable states $\rho$, if the realigned matrix $\rho^R$ has $k$ non-zero singular values $\sigma_1,\ \sigma_2,\ \cdots,\ \sigma_k$, the following inequality holds,
\begin{equation}\label{agg}
L_6:\ k(k-1) D_k^{1 / k}+T_1-1 \leqslant 0,
\end{equation}
where $D_k=\prod_{i=1}^k \sigma_i^2$ and $T_1=\operatorname{Tr}\left[(\rho^R)^\dagger \rho^R\right]$. Recently, Wang $et\ al.$ \cite{Wang_2024} generalized the method to entanglement criterion based on moments of the partial transpose.

Since the partial transposition operation is a positive but not completely positive map, it is not physically realizable and cannot be implemented experimentally. However, the related moments can be measured experimentally \cite{PhysRevLett.125.200501}. Motivated by Aggarwal $et\ al.$ \cite{PhysRevA.109.012404} and Wang $et\ al.$ \cite{Wang_2024}, in this paper we construct experimentally friendly separability criteria based on any positive but not completely positive maps.

\section{Experimentally implemental separability criteria}

Peres \cite{PhysRevLett.77.1413} introduced a powerful necessary condition for separable states, known as the positive partial transpose (PPT) criterion. If a state $\rho$ is separable, the partial transposed matrix $\rho^{T_B}$, with matrix elements given by $\la m|\la\mu |\rho^{T_B}| n\ra|\nu\ra \equiv \la m|\la\nu |\rho| n\ra|\mu\ra$, is positive semi-definite, $i.e.$, $\rho^{T_B}$ is still a quantum state. The partial transpose operation $T_B$ has an interpretation as a partial time reversal \cite{PhysRevA.58.826}. This criterion is both necessary and sufficient for $2\otimes 2$ and $2\otimes 3$ quantum systems \cite{HORODECKI19961}. Specifically, a two-qubit state is separable if and only if  $\det (\rho^{T_B})\geqslant 0$ \cite{PhysRevA.71.044301,PhysRevA.77.030301}.
Generally, for any positive map $\Lambda$, one has $(I \otimes \Lambda)\left(\rho\right) \geqslant 0$ for any separable states $\rho$. A state $\rho$ is separable if and only if  $(I \otimes \Lambda)\left(\rho\right) \geqslant 0$ for all positive maps $\Lambda$ \cite{HORODECKI19961}.

The characteristic equation of a Hermitian operator $M$ is given by
$\operatorname{det}\left(\lambda I-M\right)=0$, i.e., $\lambda^{d}+D_1 \lambda^{d-1}+D_2 \lambda^{d-2}+\cdots+D_{d}=0$, where $I$ is the identity operator. According to the Newton polynomials and the Faddeev-LeVerrier algorithm for the characteristic polynomial and traces of powers of a matrix \cite{Surles1984,Faddeev1972,ZEILBERGER1984319}, the coefficients $\left\{D_i\right\}_{i=1}^{d}$ are given by the moments of $M$,
\begin{equation}\label{coeff_chara}
D_i=(-1)^i \frac{1}{i !}\left|\begin{array}{cccccccc}
T_1 & T_2 & T_3 & . & . & . & \cdots & T_i \\
1 & T_1 & T_2 & T_3 & . & . & \cdots & T_{i-1} \\
0 & 2 & T_1 & T_2 & T_3 & . & \cdots & T_{i-2} \\
0 & 0 & 3 & T_1 & T_2 & T_3 & \cdots & T_{i-3} \\
. & . & . & . & . & . & \cdots & . \\
. & . & . & . & . & . & \cdots & . \\
. & . & . & . & . & . & \ddots & . \\
0 & 0 & 0 & 0 & 0 & 0 & \cdots & T_1
\end{array}\right|
\end{equation}
for $i=1,\ \cdots,\ d$, where $D_{d}=\operatorname{det}\left(M\right)$, $T_k=\operatorname{Tr}$ $\left[M^k\right]$ denotes the $k$th moment of $M$. For example, $D_1=-T_1$,
$D_2=\frac{1}{2}\left(T_1^2-T_2\right)$, $D_3=-\frac{1}{6}\left(T_1^3-3 T_1 T_2+2 T_3\right)$ and $D_4=\frac{1}{24}\left(T_1^4 - 6T_1^2T_2 + 8T_3T_1 + 3T_2^2 - 6T_4\right)$. In particular,
$D_r=\prod_{i=1}^r \lambda_i$, where $\lambda_i$, $i=1, \cdots, r$, are all the non-zero eigenvalues of $M$.
\begin{theorem}
If a bipartite state $\rho$ is separable, then for any positive map $\Lambda$, the coefficients of the characteristic polynomial of the matrix $(I \otimes \Lambda)(\rho)$ satisfy
\begin{equation}\label{coeff_post}
\mathrm{sgn}D_i=\left\{\begin{array}{ll}
-1, &i\leqslant r,~ i\ is\ odd,  \\
0, &i> r,\\
1, &i\leqslant r,~ i\ is\ even.
\end{array}\right.
\end{equation}
All violations of (\ref{coeff_post}) imply that the state $\rho$ is entangled.
\end{theorem}

The proof of the theorem follows from the fact that for any separable state $\rho$, $(I \otimes \Lambda)(\rho)$ is positive semi-definite. In other words, all the eigenvalues of $(I \otimes \Lambda)(\rho)$ are non-negative for separable quantum states, which implies that the coefficients of characteristic polynomial of $(I \otimes \Lambda)(\rho)$ satisfy the inequalities (\ref{coeff_post}).

For two qubit quantum states, since the partial transposed state of any entangled state is full ranked and has only one negative eigenvalue \cite{PhysRevA.58.826,FrankVerstraete_2001},
the PPT criterion is equivalent to $D_4^{PPT}< 0$, where $D_4^{PPT}$ denotes the fourth coefficient of the characteristic polynomial of $\rho^{T_B}$.

Let us consider the following PPT state $\rho \in \mathcal{B}\left(\mathcal{H}^3 \otimes \mathcal{H}^3\right)$,
$$
\rho=\frac{1}{3(3+a)}\left(\begin{array}{ccccccccc}
1 & 0 & 0 & 0 & 1 & 0 & 0 & 0 & 1 \\
0 & a & 0 & 1 & 0 & 0 & 0 & 0 & 0 \\
0 & 0 & 2 & 0 & 0 & 0 & 1 & 0 & 0 \\
0 & 1 & 0 & 2 & 0 & 0 & 0 & 0 & 0 \\
1 & 0 & 0 & 0 & 1 & 0 & 0 & 0 & 1 \\
0 & 0 & 0 & 0 & 0 & a & 0 & 1 & 0 \\
0 & 0 & 1 & 0 & 0 & 0 & a & 0 & 0 \\
0 & 0 & 0 & 0 & 0 & 1 & 0 & 2 & 0 \\
1 & 0 & 0 & 0 & 1 & 0 & 0 & 0 & 1
\end{array}\right),\ a \geqslant \frac{1}{2}.
$$

The realigned matrix of $\rho$ is given by
$$
\rho^R=
\frac{1}{3(3+a)} \left(\begin{array}{ccccccccc}
1 & 0 & 0 & 0 & a & 0 & 0 & 0 & 2 \\
0 & 1 & 0 & 1 & 0 & 0 & 0 & 0 & 0 \\
0 & 0 & 1 & 0 & 0 & 0 & 1 & 0 & 0 \\
0 & 1 & 0 & 1 & 0 & 0 & 0 & 0 & 0 \\
2 & 0 & 0 & 0 & 1 & 0 & 0 & 0 & a \\
0 & 0 & 0 & 0 & 0 & 1 & 0 & 1 & 0 \\
0 & 0 & 1 & 0 & 0 & 0 & 1 & 0 & 0 \\
0 & 0 & 0 & 0 & 0 & 1 & 0 & 1 & 0 \\
a & 0 & 0 & 0 & 2 & 0 & 0 & 0 & 1
\end{array}\right).
$$
From the realignment criterion that the trace norm $||\rho^R||_1=\operatorname{Tr}(\sqrt{(\rho^R)^\dagger\rho^R})\leqslant 1$ for all separable states \cite{rudolph2005further,Chen2002Quantum}, the state is entangled for $a<1$.

Let us consider an indecomposable map $\Gamma$ given in Ref.~\cite{Hou_2010}, which maps a matrix $A=\left(a_{i j}\right)$ to the matrix
$$
\Gamma(A)=\frac{1}{2}\left(\begin{array}{ccc}
a_{11}+a_{22} & -a_{12} & -a_{13} \\
-a_{21} & a_{22}+a_{33} & -a_{23} \\
-a_{31} & -a_{32} & a_{33}+a_{11}
\end{array}\right).
$$
Thus,
$$
\left(I \otimes \Gamma\right)\left(\rho\right)=
\frac{1}{6(3+a)} \left(\begin{array}{ccccccccc}
1+a & 0 & 0 & 0 & -1 & 0 & 0 & 0 & -1 \\
0 & a+2 & 0 & -1 & 0 & 0 & 0 & 0 & 0 \\
0 & 0 & 2+1 & 0 & 0 & 0 & -1 & 0 & 0 \\
0 & -1 & 0 & 2+1 & 0 & 0 & 0 & 0 & 0 \\
-1 & 0 & 0 & 0 & 1+a & 0 & 0 & 0 & -1 \\
0 & 0 & 0 & 0 & 0 & a+2 & 0 & -1 & 0 \\
0 & 0 & -1 & 0 & 0 & 0 & a+2 & 0 & 0 \\
0 & 0 & 0 & 0 & 0 & -1 & 0 & 2+1 & 0 \\
-1 & 0 & 0 & 0 & -1 & 0 & 0 & 0 & 1+a
\end{array}\right).
$$
It is verified that with respect to the eigenvector $(1,0,0,0,1,0,0,0,1)^T$ of $(I \otimes \Gamma)\left(\rho\right)$, one has the eigenvalue $\frac{a-1}{18+6a}$, which is negative for $a<1$. This ensures that the PPT state $\rho$ is entangled for $a<1$.

By computation, we find that the criteria given by $L_1$, $L_2$ and $L_3$ fail to detect the entanglement of $\rho$. By $\widehat{B}_3(q)<0$ the criterion given by $L_4$ detects the entanglement for $a<1$. The realignment moment criterion given by $L_5$ in Ref.~\cite{zhang2022quantum} shows that the state is entangled for $a<0.554204$. Applying the R moments criterion given by $L_6$, one obtains that the state is entangled for $a<0.795138$.  According to our theorem 1, we get from $D_9^\Gamma >0$ that the state is entangled for $a<1$. Therefore, our criterion detects the entanglement better than the ones given by $L_1$, $L_2$, $L_3$, $L_5$ and $L_6$ in this case.

\section{Conclusion}
We have investigated entanglement certification based on positive maps. By employing the Faddeev-LeVerrier algorithm, which bridges the coefficients of characteristic polynomials and the moments, we have proposed an experimentally achievable entanglement criterion, which is both sufficient and necessary for two-qubit systems. Notably, our method identifies the presence of negative eigenvalues of an operator rather than the exact eigenvalues. We have demonstrated the effectiveness of our approach by detailed PPT states, showing that more entangled states can be detected, compared with some existing moments based entanglement criteria.

\bigskip
\noindent{\bf Acknowledgments}\, \,
This work is supported by the National Natural Science Foundation of China (NSFC) (Grant Nos.~12075159, 12171044);, the specific research fund of the Innovation Platform for Academicians
of Hainan Province, Natural Science Foundation of Hunan Province (Grant No.~2025JJ60025), Scientific Research Project of the Education Department of Hunan Province (Grant No.~24B0298) and Changsha University of Science and Technology (Grant Nos.~097000303923, 097000100917).

\bibliographystyle{apsrev4-2}
\bibliography{zqhref}

\end{document}